

\magnification 1200
\tolerance 3000

\parskip 3truemm

\def\.{\cdot}

\def\bi{\bigskip \indent}

\def\ni{\noindent}

\def\nostrocostrutto#1\over#2{\mathrel{\mathop{\kern 0pt \rlap
  {\raise.2ex\hbox{$#1$}}}
  \lower.9ex\hbox{\kern-.190em $#2$}}}
\def\lsim{\nostrocostrutto < \over \sim}
\def\gsim{\nostrocostrutto > \over \sim}

\def\bi{$\tilde B$}
\def\wi{$\tilde W_3$}

\def\w3{\tilde W_3}

\def\c{\chi}
\def\den{\Omega_{\chi} h^2}
\def\m{m_{\chi}}

\def\pb{{\bar p}}
\def\tb{{\tan \beta}}

\null
{
\nopagenumbers
\rightline{DFTT 35/94}
\rightline{August 1994}
\vskip 1truecm

\centerline{\bf AMOUNT  OF ANTIPROTONS IN COSMIC RAYS}
\centerline{\bf DUE TO HALO NEUTRALINO ANNIHILATION}

\vskip 1.0truecm
\centerline{A. BOTTINO, C. FAVERO, N. FORNENGO, G. MIGNOLA}
\baselineskip=13pt
\centerline{\it Dipartimento di Fisica Teorica, Universit\`a di Torino and}
\baselineskip=12pt
\centerline{\it INFN, Sezione di Torino, via P.Giuria 1, 10125 Torino, Italy}

\vskip 3truecm

\centerline{\bf  Abstract}

We evaluate the antiproton--to--proton flux ratio which can be
generated by neutralino--neutralino annihilation in the galactic halo,
considering the most general compositions for the relic
neutralinos and modelling the neutralino local density according to
its relic abundance. We find that in the case of mixed neutralino
compositions this $\bar p/p$ ratio is higher than for
 pure higgsinos or gauginos.
It is shown how the expected improvements in sensitivity for the new
measurements of the $\bar p/p$ may provide very useful
information, complementary to the one obtainable with other experimental
means.

\vfill
\centerline{( to appear in {\it Astroparticle Physics} )}
\bigskip
\bigskip
\bigskip

\footnote{}{
E--mail addresses: (bottino, c\_favero, fornengo, mignola)@to.infn.it}

\eject
}

\pageno=1
\ni
{\bf 1. Introduction}

The neutralino ($\chi$) is generally considered the most favorite
candidate for cold dark matter, and many possible ways for finding some
evidence for it as a relic particle in the Universe have been envisaged.
              The most straightforward mean is obviously
provided by the direct method, where the incoming of a dark matter (DM)
neutralino
would be signalled by the energy released, in an appropriate detector,
by this particle when it scatters off a nucleus of the
experimental apparatus [1--2]. The existence
of relic neutralinos could also be
inferred indirectly, by detecting some specific signals due to
neutralino--neutralino annihilations. The best chances of obtaining
indirect information of this sort would be provided by neutrino
telescopes aiming at the measurement of fluxes of high--energy
neutrinos from the Sun or
from the centre of the Earth [3--4]. In fact these macroscopic bodies could
efficiently capture and accumulate relic neutralinos, and finally emit
the neutrinos produced by the neutralino pair annihilation in their
interior. To be sensitive to this measurement the neutrino
telescopes have to be of very large area ($>~ 10^5~ {\rm m^2}$)[4]. Other
indirect methods consist in the detection of signals due to
$\chi$--$\chi$ annihilation occurring in our galactic halo [5].
Theoretical predictions for the magnitude of these signals have been
performed in the case of the gamma--line [6--8], of a diffuse gamma--ray
flux [7--9], of a gamma--ray flux from the center of the Galaxy [10,8]
 and of $ \bar p$, $e^+$ components in the cosmic rays [11--14].

The issue at stake, $\it i.e.$ the detection of a DM particle in our
Universe, is so important that every method of detection is worth to be
pursued.
This is even more true, since the odds of the various experimental means
depend on the actual nature (composition and mass) of the neutralino. In the
present paper we concentrate on the evaluation of the amount of
antiprotons in cosmic rays due to the $\chi$--$\chi$ annihilation in our halo.
This problem has been previously addressed by other authors only for
neutralinos of pure compositions: either higgsinos or gauginos, and
for assigned values of the neutralino relic abundance [11--14].
At variance with
these previous investigations we consider here neutralinos of the most
general compositions and we model the neutralino local density according
to its calculated relic abundance. Our results show how sensitive
may be the antiproton--to--proton flux ratio, $ \bar p/p$, to the nature
of the neutralino.
We find that the experimental capabilities are not far from the
possibility of a fruitful search for neutralino dark matter
by measuring the $\bar p/p$ ratio. The chances of obtaining
 positive results in
these experiments depend on the actual nature  of the neutralino,
these chances being higher for a $\chi$ of mixed composition
than for the exactly pure configurations.

 \ni {\bf 2 The $\bar p$ flux}

The differential rate (per unit volume) for production
of $ \bar p$'s from $\chi$--$\chi$ annihilation is given by

$$ {dS \over {dE_{\bar p}}} = <\sigma v> f(E_{\bar p})
\left ( {{\bar {\rho}_{\chi}} \over {m_\chi}}\right )^2  \eqno(1)$$

\noindent
where $E_ {\bar p}$ is the $ \bar p$ energy, $\sigma$ is the $\chi$--$\chi$
annihilation cross section, $v$  the neutralino velocity in the halo,
and
$\bar \rho_{\chi}$ is an averaged neutralino density to be discussed
later; $f(E_{\bar p})$ denotes the $\bar p$ differential spectrum

$$f(E_{\bar p}) \equiv {1 \over \sigma} {{d \sigma (\chi \chi \rightarrow
 \bar p + X)} \over {dE_{\bar p}}} =
\sum_{F,f}
B^{(F)}_{\c f}{dN^{f}_{\bar p}\over dE_{\bar p}}  \eqno(2)$$

\noindent
where $F$ denotes the
$\c$--$\c$ annihilation final states,
$B^{(F)}_{\c f}$ is the branching ratio into
quarks or gluons in the channel $F$;
$dN^f_{\bar p}/dE_{\bar p}$ is the differential distribution
of the antiprotons generated by
hadronization of quarks (with the exception of
the top quark) and of gluons.

{}From Eq.(1) the standard formula for the interstellar $\bar p$ flux,
due to $\chi$--$\chi$ annihilation, follows

$$ \Phi_{\bar p}(E_{\bar p}) =
{ 1 \over { 4 \pi}} <\sigma v> f(E_{\bar p})
\left ( {{\bar {\rho}_{\chi}} \over {m_\chi}}
\right )^2 v_{\bar p} \tau _{\bar p} \eqno(3)$$

\ni where $v_{\bar p}$ and  $\tau_{\bar p}$ are the $\bar p$ velocity and
confinement time, respectively.

Of the various quantities that enter in Eq.(3) the first two,
$<\sigma v>$ and $f(E_{\bar p})$, depend on the neutralino properties,
whereas the last two, $v_{\bar p}$ and $\tau _{\bar p}$, are only
related to the propagation of $\bar p$'s inside the galactic halo.
The quantity $\bar {\rho}_{\chi}$ depends on
the intrinsic properties of $\chi$, but also on the features
of the $\bar p$ propagation, since the average has to be taken over
the region where the $\bar p$'s diffuse.  In what follows we illustrate
how we evaluated all these quantities.

\ni {\bf 2.1 The $\chi$--$\chi$ annihilation cross section}

In the previous works [11--14] $<\sigma v>$ has been estimated by assigning
an arbitrary value to the neutralino relic abundance
$\Omega_{\chi} h^2$ and by exploiting  a standard expression relating
the relic abundance to the $\chi$--$\chi$
annihilation cross section integrated from the freeze--out temperature
to the present temperature [15]. Instead, we calculated $<\sigma v>$
in the
framework of a particle model, the Minimal Supersymmetric Standard Model
(MSSM). For the evaluation of $\sigma$ we have included all the relevant
diagrams at the tree--level, $\it i.e.$ Higgs--exchange
diagrams and Z--exchange
diagrams in the s--channel and diagrams with exchanges of squarks,
neutralinos and charginos in the t--channel,
with all the final states which may contribute to the production of
$\bar p$'s. We have also included the one--loop diagrams which generate a
two--gluon final state [16], since, as shown in Ref.[14], this contribution is
important for $\bar p$ production by $\chi$--$\chi$ annihilation in case of
neutralinos with gaugino-dominance and with a mass
$m_{\chi}$ lighter than the top mass.

Our calculations have been performed for neutralinos of the most
general composition in terms of the photino, zino and higgsino fields

$$\eqalign{\chi=a_1\tilde\gamma + a_2\tilde Z+a_3\tilde H_1^0
+a_4\tilde H_2^0}~.\eqno(4)$$

\ni $\tilde \gamma$ and $\tilde Z$ are related to the
U(1) and SU(2) neutral gauginos, \bi \ and \wi, by the standard rotation in
terms of the Weinberg angle. For other details about the
theoretical framework adopted here for our calculations we refer to our
previous paper [17]. We only remind that the neutralino sector is usually
described in terms of three parameters: $M_2$, $\mu$ and tan$\beta$.
In the present paper for tan$\beta$
we consider two representative values tan$\beta$= 2, 8 whereas the
other two parameters are varied in the ranges: 10 GeV $\leq M_2 \leq$ 6 TeV,
10 GeV $\leq |\mu| \leq$ 3 TeV.

We
introduce a composition parameter
$P = {a_1}^2 + {a_2}^2$ which gives the gaugino fractional weight.
Almost everywhere in the parameter space, at fixed
$\tan \beta$, there is a one--to--one
correspondence: $(M_2, \mu) \leftrightarrow (P, m_{\chi})$;
thus in what follows we replace the two parameters $(M_2, \mu)$ with
($P$, $m_{\chi}$), since these variables  have a more
direct physical meaning for the neutralino.

As for the unknown masses ({\it i.e.} masses for Higgs bosons,
Susy particles and top quark), it has to be specified that most of the
results of our analysis presented here refer to the following set of
values: $m_t = 170 ~{\rm GeV}$, $m_h = 50 ~{\rm GeV}$,
$m_{\tilde f} = 1.2~ m_\c$,  when $m_\c > 45$ GeV,
$m_{\tilde f} = 45$ GeV
otherwise, except for the mass
of the top scalar partners (the only ones relevant to radiative corrections)
which has been taken $\tilde m = 1$ TeV.
Only in a few cases, other choices, to be
defined later, have been adopted for some of these
masses.

\ni {\bf 2.2  The $\bar p$ differential spectrum}

Let us consider now the $\bar p$ differential
distribution $f(E_{\bar p})$. The branching ratios $B^{(F)}_{\c f}$
have been calculated for all annihilation final states which may produce $\bar
p$'s. These states fall into two categories: i) direct production of quarks and
gluons, ii) generation of quarks through intermediate production of
Higgs bosons, gauge bosons and $t$ quark. In order to obtain the distributions
$dN^f_{\bar p}/dE_{\bar p}$ the hadronization of quarks
(with the exception of the top quark) and gluons has been
evaluated by using the Monte Carlo code Jetset 7.2 [18,19].

The ensuing differential distributions show the usual
scaling properties in the variable
$x=E_{\bar p} / (E_{\rm cm}/2)$, where $E_{\rm cm}$ is the center of mass
energy [20,13].
However, sizeable violations of this scaling behavior occur for small
values of $x$ ($x \lsim 0.1$) [20]. Since this is precisely the energy range
of interest for the process under discussion, instead of using these scaling
properties, we have evaluated the differential distributions at
four different values:  $E_{\rm cm}/2 =$ 15, 30, 60, 120 GeV.
The differential distributions at different $E_{\rm cm}$,
have been obtained by interpolation.
A convenient analytic expression that fits our numerical spectra is
provided by the following form

$$
{{dN^f_{\bar p}} \over {dx}} = p_1 |w|^{\displaystyle{p_2}}~
e^{\displaystyle{-{1\over 2} {\left({w-p_3}
\over p_4\right)^2}}} \eqno(5)$$

\noindent
where $w={\rm Log}(x_c)$, $x_c=T_{\bar p} / (E_{\rm cm}/2)$ and
$T_{\bar p}$ is the $\bar p$ kinetic energy.
The parameters $p_i$'s are given in
Table 1 for the value $ (E_{\rm cm}/2) = 30~{\rm GeV}$.

As an example,  our results for the case of the differential distribution
due to the $b$ quark fragmentation are shown in Fig.1  in
terms of the variable $x_c$. Our spectra are numerically in agreement
with the ones of Ref.[13]
for the cases considered in that paper, {\it i.e.}
for the fragmentations of the $c$ and $b$ quarks.

\ni {\bf 2.3  Propagation of the $\bar p$'s in the halo}

Let us turn now to the evaluation of the quantities that depend on
properties of the $\bar p$ propagation in the halo.
In the diffusion model [21] for propagation of cosmic rays in our Galaxy
the confinement time $\tau_{\bar p}$ may be evaluated from the
formula

$$\tau_\pb={{h_h^2} \over {3D}} {{1- {1 \over 2}(h_g/h_h)^2+
{1 \over 8} (h_g/h_h)^3} \over {1- {1 \over 2}(h_g/h_h)}} \simeq
 {{h_h^2} \over {3D}} \eqno(6)$$

\ni where $h_g$ and $h_h$ are the half--thicknesses of the disk and of the
confinement region, respectively, and $D$ is the diffusion coefficient.
The approximation involved in Eq.(6) is justified since
$h_g = 100-150~{\rm pc}$ and $h_h \gsim 700~{\rm pc}$. By varying
$h_h$ over the range $1~{\rm Kpc} \leq h_h \leq 10~{\rm Kpc}$ and by
using for $D$ the numerical values of Ref.[21] (the ratio D/$h_h$ is rather
stable with values in the range
$D/h_h \simeq (1.7-3.3) \cdot 10^6~ {\rm cm}~ {\rm s}^{-1}$),
one obtains for the $\bar p$ confinement time

$$ 0.3 \cdot 10^{15}~{\rm s} \lsim \tau_\pb \lsim 6 \cdot 10^{15}~
{\rm s}\eqno(7) $$

\ni in agreement with previous evaluations [12,13]. In the following,
for definiteness,  we adopt  the
value $\tau_{\bar p} = 3 \cdot 10^{15} {\rm s}$.

Let us discuss now the quantity $\bar {\rho}_{\chi}$ appearing in
Eq.(3). This represents the average value of the neutralino density over
a sphere centered at the Earth position, with a radius of the order of $h_h$.
It is then straightforward to relate $\bar {\rho}_{\chi}$ to the local
(solar neighborhood) neutralino density $\rho_{\chi,\rm loc}$. Assuming for
the DM halo the usual distribution
$\rho(r) = \rho_{\rm loc} (1+r_\odot^2/a^2)/(1+r^2/a^2)$, where $r_\odot \simeq
8.5~{\rm Kpc}$ and $\rho_{\rm loc} \simeq 0.3~{\rm GeV}~{\rm cm}^{-3}$
one finds in general

$$ (<\rho^2>/\rho_{\rm loc}^2)^{1/2} \simeq 1.8 \qquad {\rm for}~ a = 2.0~{\rm
Kpc ~~[22]} \eqno(8)$$

$$ (<\rho^2>/\rho_{\rm loc}^2)^{1/2} \simeq 0.97 \qquad {\rm for }~ a =
7.8~{\rm
Kpc ~~[23].} \eqno(9)$$

\ni Here, to be conservative, we adopt an estimate close to the one in
Eq.(9), $\bar {\rho} \simeq \rho_{\rm loc}$, which entails a similar
relationship for the fraction of DM due to neutralinos, i.e.
${\bar \rho_{\chi}} \simeq \rho_{\chi,\rm loc}$.

\ni {\bf 2.4 Evaluation of $\rho_{\chi,\rm loc}$}

In order to assign a value to $\rho_{\chi,\rm loc}$, for any point of the
parameter space (specified by a set of values for $M_2$, $\mu$ and
$\tan\beta$), we have evaluated the neutralino relic abundance
$\Omega_{\chi} h^2$ and we have then determined the value
$R \equiv {\rm min} \{ 1, \, (\den)/
(\Omega h^2)_{\rm min} \}$,
where $(\Omega h^2)_{\rm min}$ is defined as the minimum value of
($\Omega h^2)$ compatible with
observations. Typically, taking $\Omega \gsim 0.1$ and $h \simeq 0.5$,
one has $(\Omega h^2)_{\rm min} \simeq 0.03$; this is in fact
the value employed here for $(\Omega h^2)_{\rm min}$.
Then  $\rho_{\chi,\rm loc}$ has been determined by the
standard rescaling procedure


$$ \rho_{\chi,\rm loc} = R \, \rho_{\rm loc}\, . \eqno(10)$$

The method adopted for the evaluation of $\den$ is the one illustrated
in Ref.[17]. We only remind here that $\Omega_{\chi} h^2$ is inversely
proportional to the averaged annihilation cross section $<\sigma v>$,
integrated from the freeze--out temperature to the present one. Thus it
follows that when, in a given point of the parameter space, $\sigma$
is large, in that point $\Omega_{\chi} h^2$ is depleted.

\ni {\bf 2.5 Modulation of the $\bar p$ flux}

The $\bar p$ flux, measured at Earth, may be obtained from the
interstellar $\bar p$ flux  given by Eq.(3) by taking into account
the modulation due to the solar wind. This effect is calculated here by
employing the procedure discussed in Ref.s [24,25] and also utilized in Ref.s
[11--14]. Denoting by $E$ and $T$ the $\pb$ total and kinetic
energies and by $k$ the $\bar p$ momentum measurable at Earth,
the solar--modulated $\pb$ flux is given by

$$ \Phi_\pb(E) = {{ E^2 - m_p^2} \over { E_\pb^2 - m_p^2}} \Phi_\pb(E_\pb)~~.
\eqno(11) $$

\ni $E$ and $E_\pb$ are related by

$$\eqalignno{ E_\pb &= k_c \ln{{k+E} \over {k_c+E_c}} +E_c+ \Delta E \qquad
{\rm for} ~k < k_c \cr
E_\pb &= E+ \Delta E \qquad \qquad \qquad \qquad
\quad {\rm for}~k \geq k_c &(12) \cr }$$

\ni The numerical values for $k_c$ and $\Delta E$ depend on the phase of
the 11 year solar cycle. In what follows only the antiproton--to--proton flux
ratio $\bar p /p$, rather than the individual $\bar p$ and $p$ fluxes, will be
considered.
Since these two fluxes are
modulated in the same way, $\bar p /p$ is
quite insensitive to the specific values adopted for
$k_c$ and $\Delta E$. To be definite we use in the following the
values corresponding to a period of minimum
solar activity:
$k_c = 1.015$ GeV and $\Delta E = 495$ MeV.

\ni {\bf 3 Results and conclusions}

{}From Eq.s(3) and (10) we see that, at fixed $m_{\chi}$, the
flux $\Phi_{\bar p}$ depends on
the neutralino properties mainly through the quantity
$<\sigma v> R^2$ (some further, less effective
dependence, is introduced in $\Phi_{\bar p}$ through $f(E_{\bar p})$).
Thus we start the presentation of our results by discussing
the behavior of ${<\sigma v> R^2}$ as we move in the neutralino
parameter space.
In Fig.2 we display $<\sigma v>$, $R^2$
and  their product $<\sigma v> R^2$
as  functions of $P$, for $m_\chi =$ 30 GeV,
$\tan \beta = 2, 8$ and for the set of mass parameters ($m_h$ and
$m_{\bar f}$) previously stated.
In this figure we note a number of interesting  properties: i) the
$\chi$--$\chi$ annihilation cross section is maximal for large
higgsino--gaugino mixing ($P \simeq 0.5$) (this feature is expected,
since the small value used here for $m_h$
makes the Higgs--exchange amplitude of the annihilation cross section
very large and dominant over the other amplitudes; this effect
is particularly important when $\chi$ is largely mixed); ii) the
neutralino relic abundance $\Omega_{\chi} h^2$, and then also $R^2$, are
strongly
depleted except for rather pure compositions,
since the annihilation cross section is very efficient unless
$P \simeq 0$ or $P \simeq 1$ (for the reason mentioned at point i));
iii) from the plot of  $<\sigma v> R^2$ one sees that the combination
of the two previous properties makes the neutralinos
with a mixed composition ( $P \simeq 10^{-3}- 10^{-2}$
in the present representative points) as the most favourite ones for
detection by the method under
discussion; iv) exactly pure compositions ($P = 0,1$)
provide rather low values for  $<\sigma v> R^2$ and are in general not
indicative
of what the real situation could be when some mixing is turned on.

In Fig.3 we show how the quantity $<\sigma v> R^2$ depends on the values
assigned to the masses $m_h$ and $m_{\tilde f}$. The dashed lines
correspond to the values: i) $m_h = 50$ GeV, $m_{\tilde f} = 150$ GeV in
Fig.3a, ii) $m_h = 120$ GeV, $m_{\tilde f} = 45$ GeV in Fig.3b,
iii) $m_h = 120$ GeV, $m_{\tilde f} = 150$ GeV in Fig.3c. The solid line
always denotes the case where $m_h = 50$ GeV, $m_{\tilde f} = 45$ GeV;
this line, already shown in Fig.2, is displayed here only for sake of
comparison. In Fig.3a we see that a marked difference between the two
curves shows up  at a mixed composition
with gaugino dominance. This is expected
since in this case
sfermion--exchange diagrams are important; then, for the
gaugino--dominated neutralinos, an increase of
$m_{\tilde f}$ entails a sizeable suppression in the $\chi$--$\chi$
annihilation cross section with a subsequent enhancement of the relic
abundance and then of the $R^2$ scaling factor.
Fig.3b shows a similar effect of enhancement; however, here the increase
in $R^2$, and then in $<\sigma v> R^2$ , occurs for the mixed
compositions, since these are very sensitive to the value employed for
$m_h$. Finally in Fig.3c we notice that, when $m_h$ and $m_{\tilde f}$
are both large, the curve has a marked peak for $P \simeq 0.5$; as a
matter of fact in this case rescaling is practically absent and
$<\sigma v> R^2 \simeq <\sigma v>$.

In Fig.4 the ratio of antiproton--to--proton integrated fluxes,
$(\bar p/p)_{\rm int}$, is displayed versus $P$ for $m_{\chi} = 30$ GeV and
tan $\beta = 2,8$. The $\bar p$ spectrum has been evaluated from Eq.(3)
and the proton spectrum is taken from Ref. [26]:
$ \Phi_p(E_ p) = 1.93 ~ v_p (E_p/{\rm GeV})^{-2.7} ~{\rm cm}^{-2}
{\rm s}^{-1} {\rm sr}^{-1} {\rm GeV}^{-1}$ ($v_p$ and $E_p$ are the
proton velocity and energy).
The modulation for both
spectra is obtained with the procedure discussed above in Sect.2.5.
Both $p$ and $\bar p$ spectra have been integrated over the ranges:
(a) 100 MeV$ \leq T \leq $ 640 MeV; (b) 640 MeV$ \leq T \leq $ 1580 MeV;
and (c) 120 MeV$ \leq T \leq $860 MeV in order to conform to the energy
windows used in the experiments of Ref.s [27,28]. The horizontal
lines in Fig.4 denote the corresponding upper bounds: $(\bar p /p)_{\rm int} =
 2.8 \cdot 10^{-5}$ (85 \% C.L.) for the interval a) [27];
$(\bar p /p)_{\rm int} = 6.1 \cdot 10^{-5}$ (85 \% C.L.)
for the interval b) [27];
$(\bar p /p)_{\rm int} = 1.8 \cdot 10^{-5}$ (85 \% C.L.)
for the interval c) [28].
It is remarkable
that the experimental sensitivities are already at the level of the
maximum of the theoretical curves (at the representative points under
discussion). This means that by improving these sensitivities one is
able to start investigating the relic neutralinos by the measurement
of $\bar p / p$.

Let us turn to the plots of the $\bar p/p$ ratio versus the
$\bar p$ kinetic energy $T$. Fig.5 shows how sensitively,
at fixed $m_{\chi}$, this ratio depends
on the neutralino composition. In this figure also the background due to
the secondary antiprotons produced in
cosmic--ray collisions in the interstellar
medium is displayed [29].

The dependence on the neutralino mass of the plots of $\bar p/p$ versus
$T$ is illustrated in Fig.6 for three values of $P$ separately [30].
In Fig.6c we notice that the ordering of the four curves for our signal
is the usual one, expected
on the basis of the dependence on $m_{\chi}^{-2}$ in Eq.(3). However,
in Fig.s 6a,6b this mass hierarchy turns out to be modified; this
effect occurs when $2\m$ is close to one of the masses of the particles
exchanged in the s--channel of the annihilation process. Then, in the
present case, this happens for $\m = 30$ GeV which is close to the
value used here for one of the neutral Higgs bosons.

It has to be pointed out that many uncertainties are involved
in the theoretical predictions for the $\bar p/p$ signal,
as it emerges from the derivation of the
$\bar p$ flux presented above.
Suffice it to mention the number of free
parameters in MSSM and the wide range in the estimate of the
confinement time $\tau_{\bar p}$.

Nevertheless, from the previous analysis it turns out that a substantial
improvement in the measurement of the $\bar p/p$ ratio [31] can make
this  kind of investigation
competitive in the future with other experimental
strategies which aim at finding a sign of neutralino DM. This
measurement shares with the measurements of other signals due to the
$\chi$--$\chi$ annihilation in the halo [6--10] the peculiar
possibility of providing information about
${\chi}$--compositions with small mixings, which have little chances to be
detected by other experimental means [32]. This point is illustrated in
Fig.7 where we give the $(\bar p/p)_{\rm int}$ ratio versus
$\m$ (7a) and versus $P$ (7b) in the form
of a scatter plot, obtained by varying the two parameters $M_2$
and $\mu$ in the ranges 10 GeV $\leq M_2 \leq$ 6 TeV,
10 GeV $\leq |\mu| \leq$ 3 TeV, at fixed $\tan\beta = 8$.
These two plots show how the exploration may proceed
inside the DM neutralino physical region, especially at small $P$ and
small $m_{\chi}$, by increasing the experimental
sensitivity in the measurement of $\bar p/p$.

The most serious problem for the detection of the neutralino signal
under discussion  is
due to the uncertainty in the
evaluation of the flux of secondary $\bar p$'s
produced by cosmic--ray collisions; this flux
plays here the role of a background. Unfortunately, the calculation of
this spectrum is affected by large uncertainties at small $\bar p$
kinetic energies, making the signal--to--background discrimination
difficult. Then a more accurate evaluation of this background
is very much needed.
However, it has to be emphasized that new stringent upper
limits in the range of small energies would anyway provide useful
constraints about the neutralino physical region.
\vfill
\eject

\ni {\bf References}

\item{[1]} For an up--to--date review about experiments for direct detection
of dark matter particles see  L.Mosca, invited talk at the XIVth
Moriond Workshop, 1994 (Proc., to appear). \hfill \break

\item{[2]} A. Bottino, V. de Alfaro, N. Fornengo, G. Mignola and
S. Scopel, {\it Astroparticle Physics} {\bf 2} (1994) 77, and references quoted
therein.\hfill \break

\item{[3]} For a general overview about the experimental capabilities
in this field see {\it Proc. of the Sixth International Workshop on
Neutrino Telescopes} (Venezia 1994, Ed. Milla Baldo Ceolin), to appear.
\hfill \break

\item{[4]} A. Bottino, N. Fornengo, G. Mignola and L. Moscoso: {\it Signals of
neutralino dark matter from Earth and Sun}, University of Torino preprint
DFTT 34/94 (July 1994), and references quoted therein.\hfill \break

\item{[5]} Signals due to non--baryonic dark matter annihilations
in the Large Magellanic Cloud have been considered in P. Gondolo,
{\it Nucl. Phys.} (Proc. Suppl.) {\bf B35} (1994) 148 (TAUP 93 Proceedings,
Ed.s C. Arpesella, E. Bellotti and A. Bottino).

\item{[6]} L. Bergstr\"om and H. Snellman, {\it Phys. Rev.} {\bf D37}
 (1988) 3737; \hfill\break
S. Rudaz, {\it Phys. Rev.} {\bf D39} (1989) 3549; \hfill \break
G.F. Giudice and K. Griest,
{\it Phys. Rev.} {\bf D40} (1989) 2549; \hfill \break
A. Bouquet, P. Salati and J. Silk, {\it Phys. Rev.} {\bf D40} (1989) 3168;
\hfill \break
L. Bergstr\"om, {\it Phys. Lett.} {\bf B225} (1989) 372; \hfill \break
S. Rudaz and F.W. Stecker, {\it Ap. J.} {\bf 368} (1991) 406; \hfill \break
V.S. Berezinsky, A. Bottino and V. de Alfaro, {\it Phys. Lett.}
{\bf B271} (1992) 122; \hfill \break
L. Bergstr\"om and J. Kaplan: {\it Gamma ray lines from TeV dark matter},
 USITP--94--03 preprint. \hfill \break

\item{[7]} S. Rudaz and F.W. Stecker,
{\it Ap. J.} {\bf 325} (1988)16; \hfill \break
F.W. Stecker and A.J.Tylka, {\it Proc. 21st ICRC} (Adelaide, 1990), vol. 1,
p.142. \hfill \break

\item{[8]} M. Urban, A. Bouquet, B. Degrange, P. Fleury,
J. Kaplan, A.L. Melchior and E.
Par{\'e}, {\it Phys. Lett.} {\bf B293} (1992) 149. \hfill \break

\item{[9]} K. Freese and J. Silk, {\it Phys. Rev.} {\bf D40} (1989) 3828;
\hfill
\break
H.--U. Bengtsson, P. Salati and J. Silk,
{\it Nucl. Phys.} {\bf B346} (1990) 129.
\hfill \break

\item{[10]} J.Silk and H. Bloemen, {\it Ap. J. Lett.} {\bf 313} (1987) L47;
\hfill \break
J.R. Ipser and P. Sikivie, {\it Phys. Rev.} {\bf D35} (1987) 3695; \hfill
\break
V.S. Berezinsky, A.V. Gurevich and K.P. Zybin, {\it Phys. Lett.} {\bf B294}
(1992) 221; \hfill \break
J. Silk and A. Stebbins {\it Ap. J.} {\bf 411} (1993) 439; \hfill \break
V.S. Berezinsky, A. Bottino and G. Mignola, {\it Phys. Lett.}
{\bf B325} (1994) 136. \hfill \break

\item{[11]} F.W. Stecker, S. Rudaz and T.F. Walsh, {\it Phys. Rev. Lett.} {\bf
55} (1985) 2622.

\item{[12]} S. Rudaz and F.W. Stecker, {Ap. J.} {\bf 325} (1988) 16.

\item{[13]} J. Ellis, R.A. Flores, K. Freese, S. Ritz, D. Seckel and J. Silk
{\it Phys. Lett.} {\bf B214} (1988) 403.

\item{[14]} G. Jungman and M. Kamionkowski, {\it Phys. Rev.} {\bf D49} (1994)
2316.

\item{[15]} J. Ellis, J.S. Hagelin, D.V. Nanopoulos, K. Olive and M. Srednicki,
{\it Nucl. Phys.} {\bf B238} (1984) 453.

\item{[16]} M. Drees, G. Jungman, M. Kamionkowski and M.M. Nojiri, {\it
Phys. Rev.} {\bf D49} (1994) 636.

\item{[17]} A. Bottino, V.de Alfaro, N. Fornengo, G. Mignola and M. Pignone,
{\it Astroparticle Physics} {\bf 2} (1994) 67.

\item{[18]} T. Sj\"ostrand, {\it Comp. Phys. Comm.} {\bf 39} (1986) 347;
{\it Comp. Phys. Comm.} {\bf 43} (1987) 367;
CERN--TH 6488/92.

\item{[19]} The fragmentation functions for $c$ and $b$ quarks used
here are the ones suggested by  C.Peterson, D. Schletter, L. Schmitt
and P.M Zerwas, {\it Phys. Rev.} {\bf D27} (1983) 105.

\item{[20]} P. M\"attig, {\it Phys. Rep.} {\bf 177} (1989) 141.

\item{[21]} V.S. Berezinsky, S.V. Bulanov, V.A. Dogiel, V.L. Ginzburg and
V.S. Ptuskin: {\it Astrophysics of Cosmic Rays} (North--Holland, 1990) ch.3.

\item{[22]} J.N.Bahcall and R.M.Soneira, {\it Ap. J. Suppl.} {\bf 44} (1980)
73.

\item{[23]} J.A.R.Caldwell and J.P.Ostriker, {\it Ap. J.} {\bf 251} (1981) 61.

\item{[24]} L.A. Fisk, {\it J. Geophys. Res.} {\bf 76} (1971) 221.

\item{[25]} J.S. Perko, {\it Astron. Astrophys.} {\bf 184} (1987) 119.

\item{[26]} M.J. Ryan, J.F. Ormes, V.K. Balasubrahmanyan, {\it Phys. Rev. Lett.
} {\bf 28} (1972) 985.

\item{[27]} S.W. Barwick et al., {\it Proc. 21st ICRC}
(Adelaide, 1990), vol. 3, p.273. \hfill \break

\item{[28]} R.E. Streitmatter et al., {\it Proc. 21st ICRC}
(Adelaide, 1990), vol. 3, p.277. \hfill \break

\item{[29]} R.J. Protheroe, {\it Ap. J.} {\bf 251} (1981) 387.

\item{[30]} In order to explore a rather wide range in the neutralino
mass, in Fig.6 we have used as a
representative $m_\c$ value for a light neutralino the value
$m_{\chi} = $15 GeV, even if, strictly speaking, this is marginally excluded
by the LEP data, which give $m_{\chi} \gsim 18$ GeV (at 90 \% C.L.),
if GUT relations are  applied (L. Roszkowski, {\it Phys. Lett.}
{\bf B252} (1990) 471; K. Hidaka, {\it Phys. Rev.} {\bf D44} (1991) 927).

\item{[31]} The Balloon--borne Experiment with a Superconducting
Solenoidal Spectrometer (BESS) is expected to reach for the
$\bar p/p$ ratio the level of $10^{-6}$ (or even $10^{-7}$, by making
use of the long duration ballooning capability) in an energy range:
120--600 MeV (K. Anraku, {\it Proc.of the 23rd International
Cosmic Rays Conference} (Calgary 1993) p.156). A new project for a
satellite experiment (PAMELA) is designed to reach the level of
$0.5 \cdot 10^{-7}$ with the energy window: 100--500 MeV
(Dr. A. Morselli, private communication).

\item{[32]}
This is due to the fact that in the case of the $\chi$--$\chi$
annihilation in the halo
the signals depend quadratically on $\rho_{\chi}$ instead of linearly as
in the cases of the direct detection and of the signals from the Earth
and the Sun.
\vfill
\eject

\ni { \bf Figure Captions}
\medskip

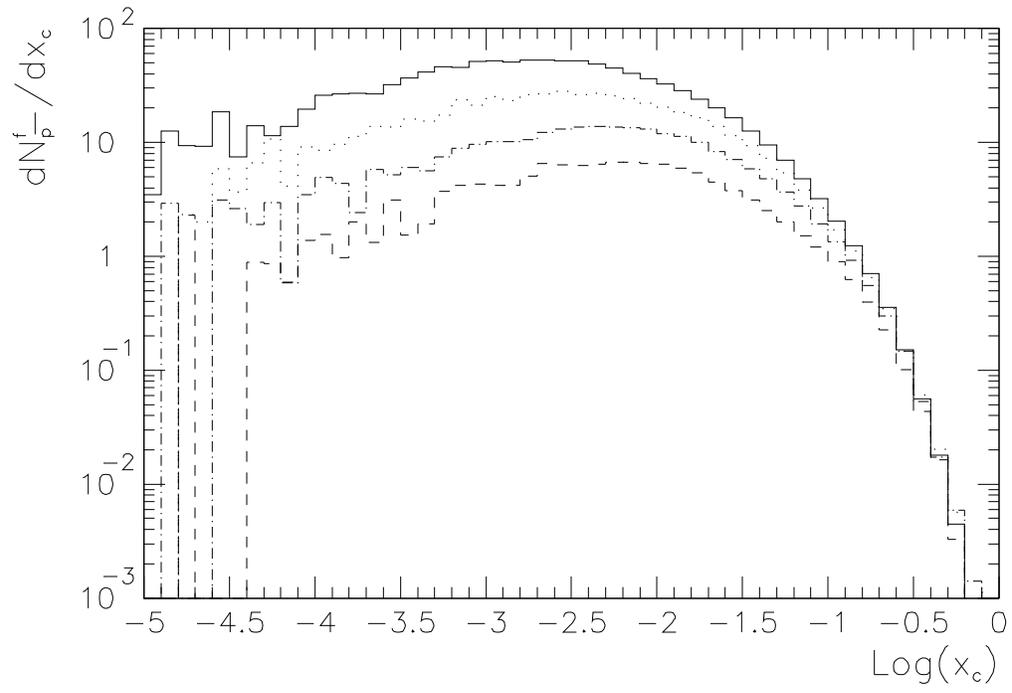
{\bf Figure 1}
$ \bar p$ differential distibution $dN^f _{\bar p} / dx_c$
from the fragmentation of a $b$ quark,
as a function of $x_c = T_{\bar p}/(E_{\rm cm} /2)$, $T_{\bar p}$ being
the antiproton kinetic energy.
The curves refer to four
different values of $E_{\rm cm}$. From top to bottom:
$E_{\rm cm}/2 =$ 120, 60, 30 and 15 GeV.

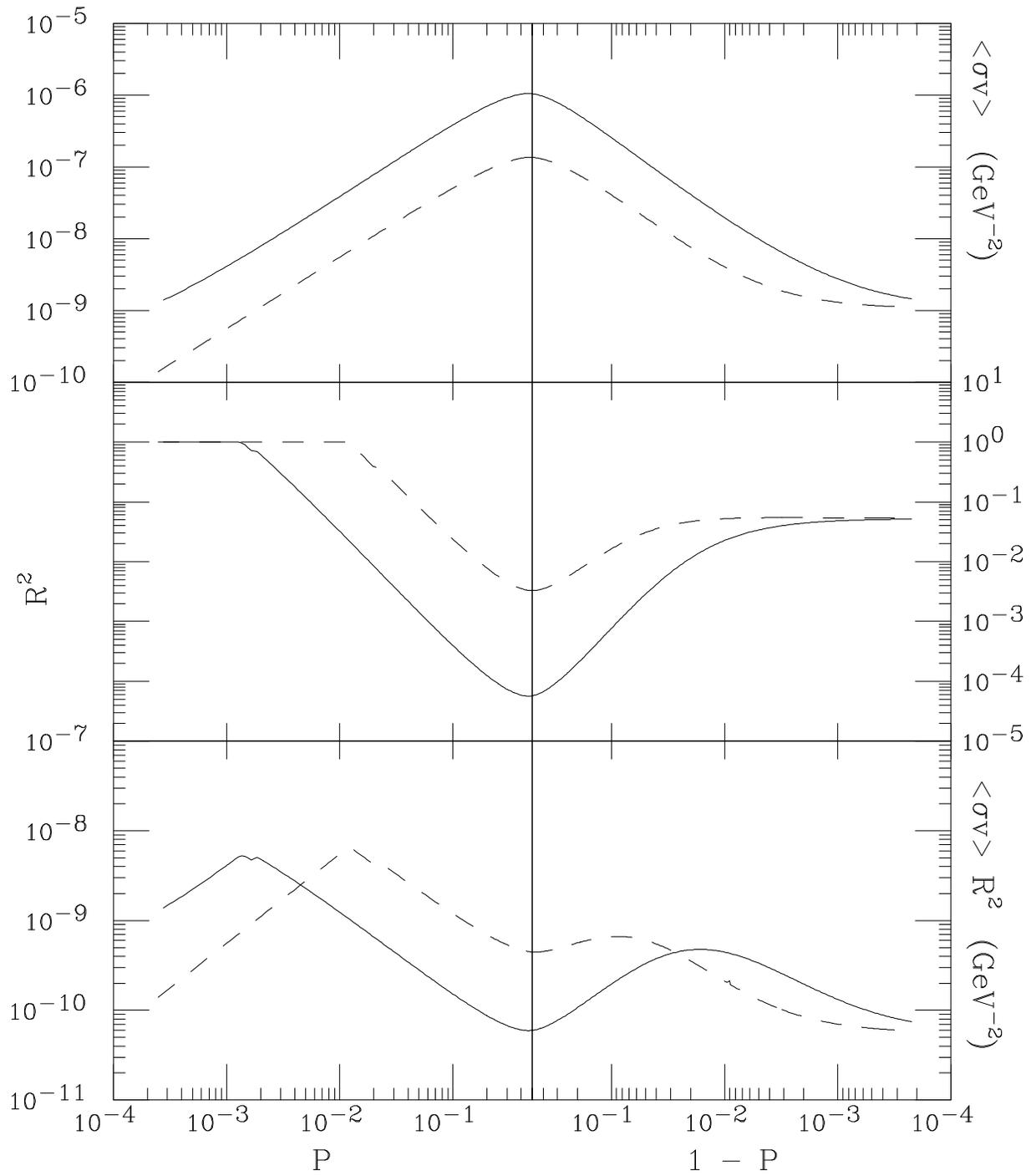
{\bf Figure 2}
$<\sigma v>$,  $R^2$
and  $<\sigma v> R^2$ as functions of
the neutralino composition at a fixed value of the neutralino mass
$m_\c=30$ GeV.  The other parameters
are: $\tb = 8$ (solid curve), $\tb = 2$ (dashed curve), $m_h = 50$ GeV,
$m_{\tilde f} = 45$ GeV.
In these plots, in order to represent the dependence of various
quantities accurately for very small mixings
({\it i.e.} for $P \sim 0, P \sim 1$), on the left part of the
horizontal axis the variable $P$ is given, whereas on the right part of
the same axis the complementary variable $1 - P$ is reported, both in
a log--scale.

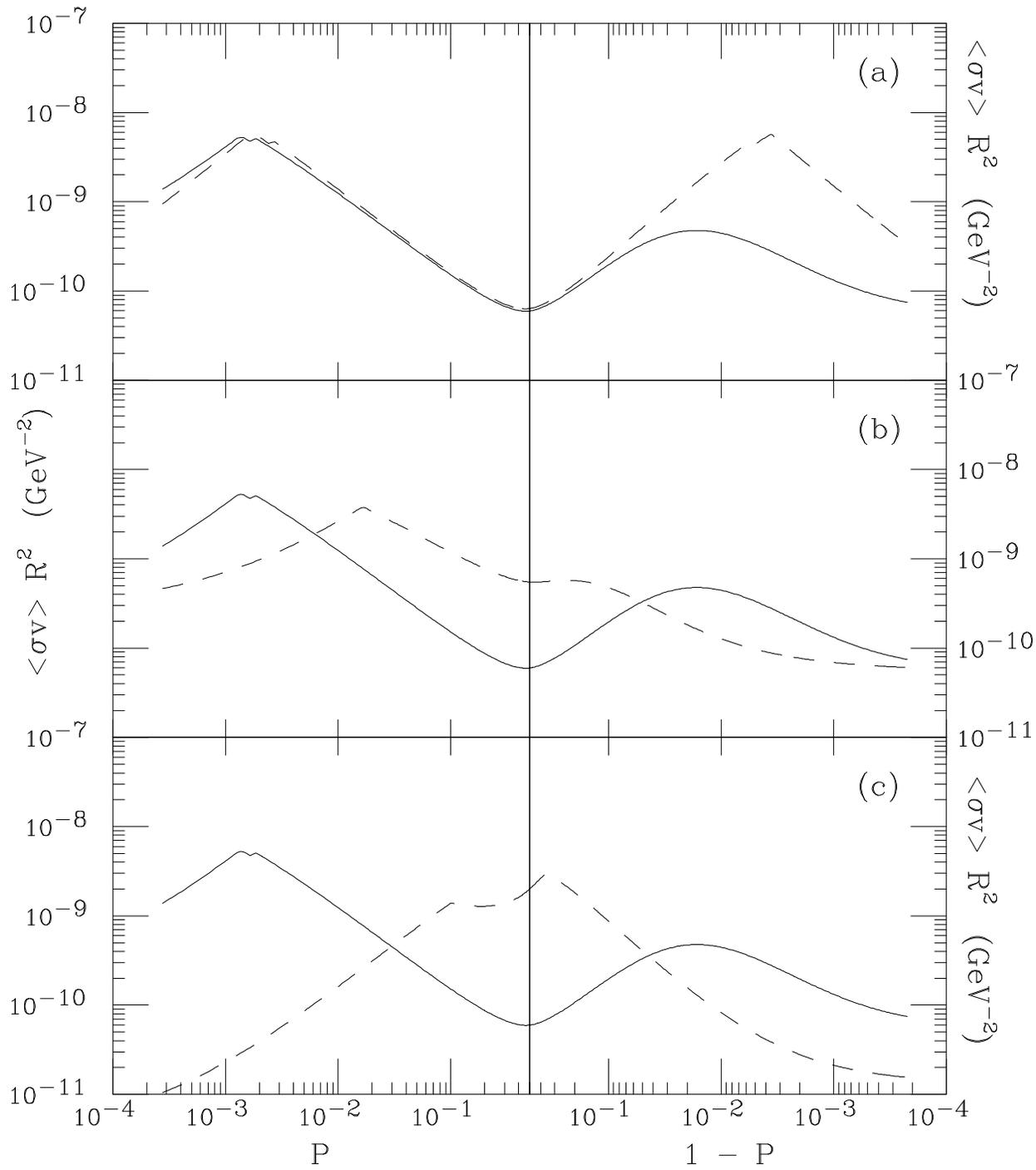
{\bf Figure 3}
$<\sigma v> R^2$ as a function
of the neutralino composition for $\tb = 8$
and for a fixed value for the neutralino mass $m_\c =30$ GeV.
The solid curve refers to
$m_h = 50$ GeV, $m_{\tilde f} = 45$ GeV.
The dashed curves denote different choices for the mass parameters:
(a) $m_h =  50$ GeV and $m_{\tilde f} = 150$ GeV;
(b) $m_h = 120$ GeV and $m_{\tilde f} =  45$ GeV;
(c) $m_h = 120$ GeV and $m_{\tilde f} = 150$ GeV.

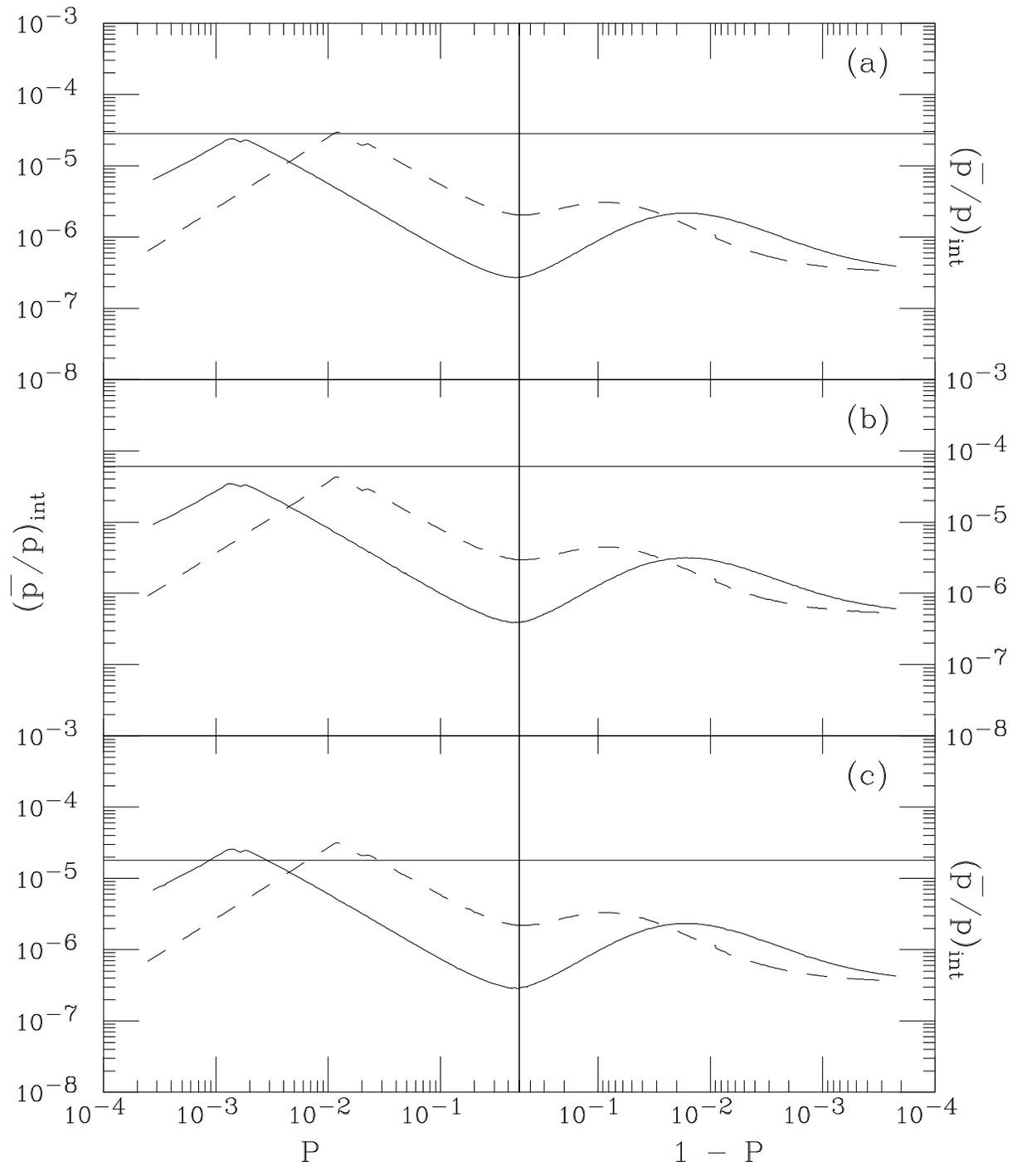
{\bf Figure 4}
Ratio of antiproton--to--proton integrated fluxes,
$({\bar p}/p)_{\rm int}$, as a function of the neutralino composition
for $\m =$ 30 GeV.
The solid curve refers to $\tb = 8$, the dashed curve to $\tb = 2$.
The mass parameters are: $m_h = 50$ GeV and $m_{\tilde f} = 45$ GeV.
The integration ranges are:
(a) 100 MeV$ \leq T \leq $ 640 MeV;
(b) 640 MeV$ \leq T \leq $ 1580 MeV;
(c) 120 MeV$ \leq T \leq $860 MeV.
The horizontal lines are the corrisponding present upper bounds:
(a) $(\bar p /p)_{\rm int} = 2.8 \cdot 10^{-5}$ (85 \% C.L.) [27];
(b) $(\bar p /p)_{\rm int} = 6.1 \cdot 10^{-5}$ (85 \% C.L.) [27];
(c) $(\bar p /p)_{\rm int} = 1.8 \cdot 10^{-5}$ (85 \% C.L.) [28].

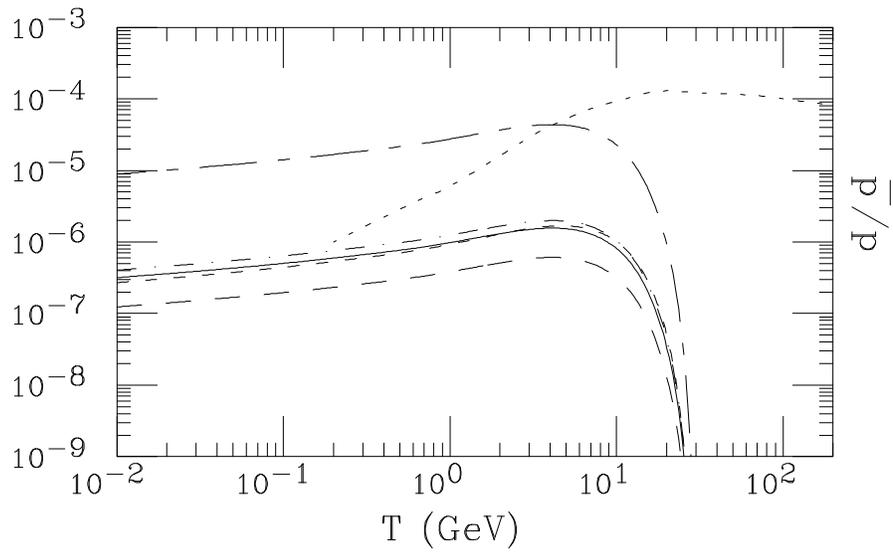
{\bf Figure 5}
Antiproton--to--proton flux ratio, ${\bar p}/p$, as a function of
the kinetic energy $T$, for $\m = 30$ GeV and for different
neutralino compositions: $P = 10^{-3}$ (long--short dashed curve);
$P = 0.1$ (solid curve); $P = 0.5$ (long dashed curve);
$P = 0.9$ (dash--dotted curve); $P = 0.999$ (short dashed curve).
The dotted curve is the background flux of secondary antiprotons,
calculated in the leaky--box model [29]. Here $\tb=8$, $m_h=50$ GeV and
$m_{\tilde f} = 45 $GeV.

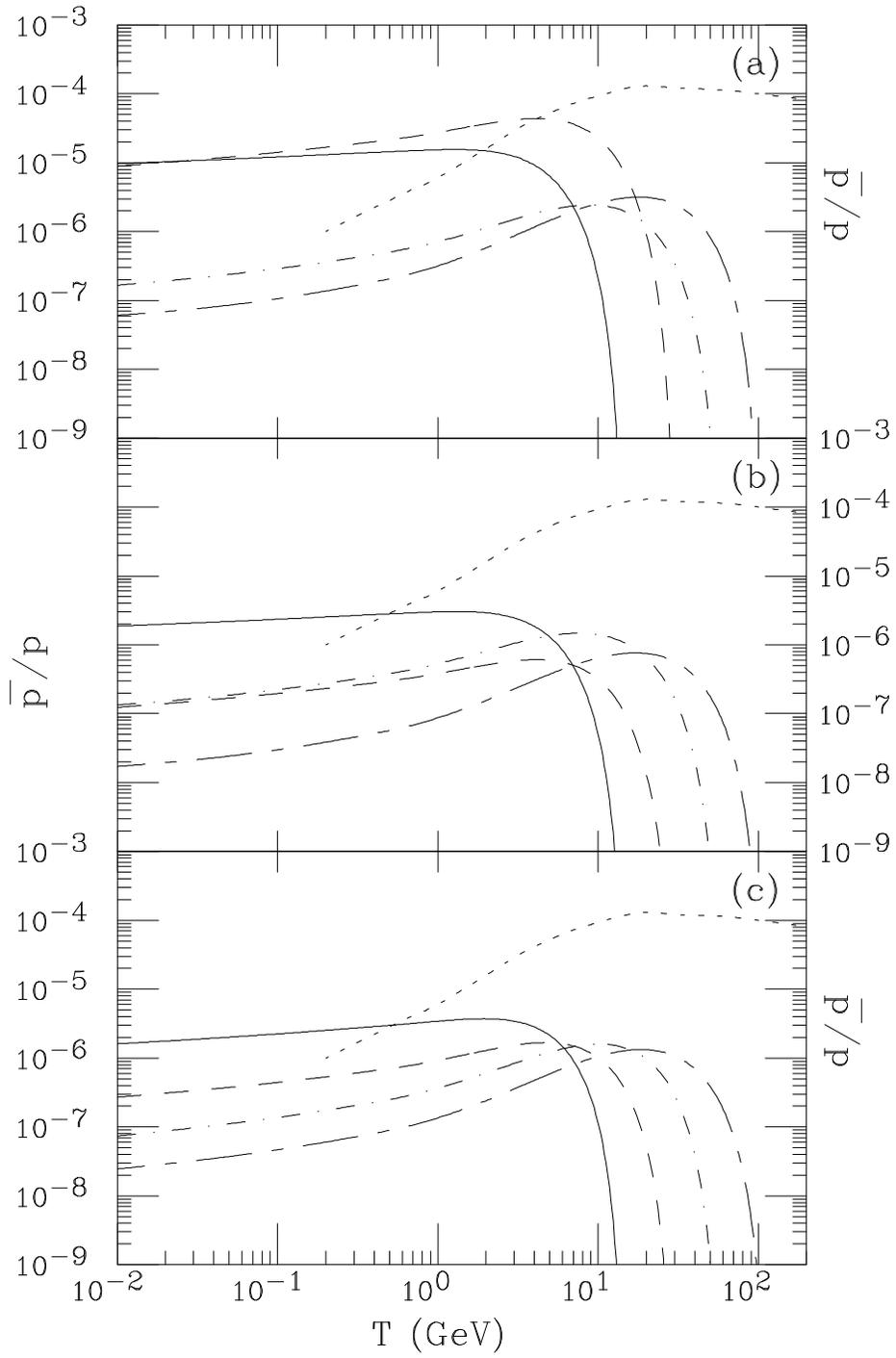
{\bf Figure 6}
Antiproton--to--proton flux ratio, ${\bar p}/p$, as a function of
the kinetic energy $T$.
Different curves refer to the following neutralino masses:
$\m = 15$ GeV (solid curve); $\m = 30$ GeV (dashed curve);
$\m = 60$ GeV (dash--dotted curve);
$\m = 120$ GeV (long--short dashed curve).
In (a) the neutralino composition is fixed at the value
$P = 10^{-3}$; in (b) $P = 0.5$; in (c) $P = 0.999$.
The dotted curve is the background flux of secondary antiprotons,
calculated in the leaky--box model [29].
Here $\tb=8$, $m_h=50$ GeV and $m_{\tilde f} = 1.2~m_\c$, when $m_\c >45$
GeV, $m_{\tilde f} = 45$ GeV otherwise.

{\bf Figure 7}
Ratio of antiproton--to--proton integrated fluxes,
$({\bar p}/p)_{\rm int}$, as a function of the neutralino mass $\m$
(a) and of the neutralino composition (b).
The figure is presented in the form of a scatter plot obtained
by varying the model parameters in the ranges:
10 GeV $\leq M_2 \leq$ 6 TeV, 10 GeV $\leq |\mu| \leq$ 3 TeV.
The integration interval is 120 MeV$ \leq T \leq $860 MeV.
The horizontal line is the corresponding upper bound
$(\bar p /p)_{\rm int} = 1.8 \cdot 10^{-5}$ (85 \% C.L.) [28].
Here $\tb=8$, $m_h=50$ GeV and $m_{\tilde f} = 1.2~m_\c$, when $m_\c >45$
GeV, $m_{\tilde f} = 45$ GeV otherwise.

\vfill
\eject

\ni {\bf Table Caption}
\medskip

{\bf Table1} Values for the $p_i$'s parameters of the distribution given in
Eq.(5) for the value $E_{\rm cm}/2=30$ GeV. Each row corresponds to a quark
flavour and to the gluon.
\medskip

$$
\vbox{ \offinterlineskip \hrule
\halign{ \vrule#&&
  \strut\quad \hfil#\hfil\quad&\vrule#\cr
height2pt&\omit&&\omit&&\omit&&\omit&&\omit&\cr
&\  \hfil&&$p_1$&&$p_2$&&
$p_3$&&$p_4$&\cr
height2pt&\omit&&\omit&&\omit&&\omit&&\omit&\cr
\noalign{\hrule}
height2pt&\omit&&\omit&&\omit&&\omit&&\omit&\cr
&$u,d$&& 1.7443&& 2.9722&&-0.62737&&-1.1193&\cr
height2pt&\omit&&\omit&&\omit&&\omit&&\omit&\cr
\noalign{\hrule}
height2pt&\omit&&\omit&&\omit&&\omit&&\omit&\cr
&$s$&& 7.5915&& 4.0079&& 1.5257&&-1.4500&\cr
height2pt&\omit&&\omit&&\omit&&\omit&&\omit&\cr
\noalign{\hrule}
height2pt&\omit&&\omit&&\omit&&\omit&&\omit&\cr
&$c$&& 36.268&& 4.8055&& 3.0725&&-1.5923&\cr
height2pt&\omit&&\omit&&\omit&&\omit&&\omit&\cr
\noalign{\hrule}
height2pt&\omit&&\omit&&\omit&&\omit&&\omit&\cr
&$b$&& 6.9350&& 5.4031&& 1.7076&&-1.3286&\cr
height2pt&\omit&&\omit&&\omit&&\omit&&\omit&\cr
\noalign{\hrule}
height2pt&\omit&&\omit&&\omit&&\omit&&\omit&\cr
&gluon&& 132.86&& 4.9732&& 3.5982&&-1.6360&\cr
height2pt&\omit&&\omit&&\omit&&\omit&&\omit&\cr}
\hrule}
$$

\bye